# Superfolded configuration induced low thermal conductivity in two-dimensional carbon allotropes revealed *via* machine learning force constant potential

Linfeng Yu[1], Kexin Dong[1], Qi Yang[1], Yi Zhang[1], Xiong Zheng[1], Huimin Wang[2], Zhenzhen Qin[3], and Guangzhao Qin[1,4,5*]

[1]*National Key Laboratory of Advanced Design and Manufacturing Technology for Vehicle, College of Mechanical and Vehicle Engineering, Hunan University, Changsha 410082, P. R. China*
[2]*Hunan Key Laboratory for Micro-Nano Energy Materials & Device and School of Physics and Optoelectronics, Xiangtan University, Xiangtan 411105, Hunan, China*
[3]*School of Physics and Microelectronics, Zhengzhou University, Zhengzhou 450001, China*
[4]*Research Institute of Hunan University in Chongqing, Chongqing 401133, China*
[5]*Greater Bay Area Institute for Innovation, Hunan University, Guangzhou 511300, Guangdong Province, China*

## 0. Abstract

Understanding the fundamental link between structure and functionalization is crucial for the design and optimization of functional materials, since different structural configurations could trigger materials to demonstrate diverse physical, chemical, and electronic properties. However, the correlation between crystal structure and thermal conductivity ($\kappa$) remains enigmatic. In this study, taking two-dimensional (2D) carbon allotropes as study cases, we utilize phonon Boltzmann transport equation (BTE) along with machine learning force constant potential to thoroughly explore the complex folding structure of pure $sp^2$ hybridized carbon materials from the perspective of crystal structure, mode-level phonon resolved thermal transport, and atomic interactions, with the goal of identifying the underlying relationship between 2D geometry and $\kappa$. We propose two potential structure evolution mechanisms for targeted thermal transport properties: *in-plane* and *out-of-plane* folding evolutions, which are generally applicable to 2D carbon allotropes. It is revealed that the folded structure produces strong symmetry breaking, and simultaneously produces exceptionally strongly suppressed phonon group velocities, strong phonon-phonon scattering, and weak phonon hydrodynamics, which ultimately lead to low $\kappa$. The insight into the folded effect of atomic structures on thermal transport deepens our understanding of the relationship between structure and functionalization, which offers straightforward guidance for designing novel nanomaterials with targeted $\kappa$, as well as propel developments in materials science and engineering.

---

[*] Author to whom all correspondence should be addressed. E-Mail: gzqin@hnu.edu.cn

## 1. Introduction

Two-dimensional (2D) carbon crystal structures have recently received an abundance of theoretical and experimental interest as an emerging material system [1–5]. They are the perfect objects for investigating structure-function interactions due to their various configurations and distinctive physical and chemical features, such as high carrier mobility, great mechanical strength, exceptional thermal stability, and large specific surface area [6–9]. Thermal conductivity ($\kappa$), a crucial indicator for describing heat transfer performance in materials, is directly influenced by the internal structure and phonon transport properties. In-depth study on the correlation of the internal structure and $\kappa$ characteristics can provide powerful guidance for the design and application of advanced thermal functional materials [10–13], thereby promoting the development of the field of thermal management.

The phonon behavior in 2D materials is controlled by structural confinement and dimensionality effects, which differs from those observed in conventional three-dimensional (3D) materials. In 2D materials, layered structure restricts atomic coupling to the *in-plane* direction, resulting in heat conduction primarily through *in-plane* phonon transport [14,15]. Compared to 3D materials, the *in-plane* vibrational modes of 2D crystals are strongly confined, leading to a higher density of phonon modes [15,16]. Furthermore, the phonon dispersion relation in 2D materials differs from that in 3D materials. While the phonon dispersion curve in 3D materials typically demonstrates a linear relationship, implying proportionality between phonon energy and momentum, the phonon dispersion curve in 2D materials follows a quadratic function shape [15,17]. This quadratic shape enhances the intensity and nonlinearity of the phonon dispersion, potentially causing phonon energy shifts to be more sensitive and abrupt within a given range of momentum. Additionally, the phonon propagation speed in momentum space may not be synchronized with the frequency shift. The unique properties enable 2D carbon to exhibit distinct heat conduction characteristics compared to conventional 3D materials, making it still difficult to understand the structure-$\kappa$ relationship in 2D crystal structures, especially in graphene derivatives and carbon nanostructures with complex topologies.

The diversity and complexity of atomic bonding types and dimensional effects make accurate estimations and general interpretations for the $\kappa$ of the numerous carbon allotropes more challenging, but they also inspire strong research interest. For instance, Choudhry et al. [22] investigated the thermal transport properties of solely $sp^2$ hybridized 2D carbon planar structures and discovered that the poor $\kappa$ is caused by the superstructure's acoustic mode folding effect. Our further study [19] has revealed that the insertion of the $sp^3$ hybrid structure may additionally trigger substantial phonon-phonon scattering, affecting the thermal transport properties. Interestingly, Giri et al. [20] discovered that in amorphous carbon materials, increasing the amount of $sp^3$ hybridization may greatly boost the contribution of propagating modes, resulting in a four-fold increase in $\kappa$, which seems to be the opposite of its role in 2D carbon materials. While $sp^3$ hybridized carbon atoms contribute an extra scattering mechanism that might boost the contribution of propagating modes and hence increase $\kappa$, amorphous carbon materials have carbon atoms organized chaotically. It also reveals that the interaction between the specific material structure and the pure $sp^2$ (or $sp^3$) hybrid structure affecting the heat transfer capability is complex, leading to the fact that the performance of $\kappa$ is affected by various architectures and bond hybridizations.

Purely $sp^2$-hybridized carbon atoms are arranged into planar structures in 2D carbon materials, where folding effects triggered by the superstructure limit the amount of heat that can be transported. The different folding shapes and configurations of carbon atoms may lead to different phonon propagation characteristics and $\kappa$. Note that the remarkable features of 2D carbon structures include not only *in-plane* structural folding but also *out-of-plane* morphological expansion, leaving a major gap in our understanding of the link between the shape of these structures and $\kappa$. A variety of *out-of-plane* structures, including carbon nanotubes [21,22], carbon nanoribbons [23,24], and graphene derivatives [25–29], are also present in the 2D carbon structure. These complex and rich *out-of-plane* designs generate various phonon propagation modes and scattering processes that have a significant impact on $\kappa$. However, precise $\kappa$ is elusive despite the rich and complex carbon configurations that enable diverse phonon transport behaviors. Complex and asymmetric configurations require computational resources to perform high-order partial derivatives of potential energy, especially the fourth-order and above interatomic force constants, leading to expensive costs for obtaining thermal transport properties, while solving the phonon Boltzmann transport equation (BTE) with first-principles calculations. As for the molecular dynamics method, it can consider higher-order phonon effects and may exhibit higher efficiencies, such as the four-phonons in c-BAs [30], it is highly dependent on the precise interatomic potential function. And the separation of various orders of phonon-phonon scattering is difficult. Hence, obtaining the $\kappa$ for complex carbon configurations is often difficult due to the expensive computational cost of higher-order force constants and the lack of interatomic potential functions. Although substantial progress has been achieved in understanding the $\kappa$ of 2D carbon structures [18–20,26,31–33], it is necessary to investigate the mechanism of different structural morphologies on the $\kappa$ to gain a more comprehensive understanding of the phonon propagation and scattering behavior in 2D carbon structures by more convenient methods. By bridging these knowledge gaps, the $\kappa$ of 2D carbon materials can be better understood and engineered to meet increasing application demands.

In this study, we exploit the phonon BTE combined with machine learning potential to investigate pure $sp^2$ hybrid carbon materials with complex folded structures. Compared with traditional first-principles and molecular dynamics methods, machine learning force constant potential methods can efficiently evaluate higher-order phonon effects, and can quickly obtain potential energy surfaces and their partial derivatives without consuming more computing resources for asymmetric complex lattices. In this way, we focus on the crystal structure, phonon properties, phonon-phonon scattering and atomic interaction levels of 2D carbon materials, with the goal to establish a link between different geometries and heat transfer properties (geometry-phonon-$\kappa$) in the 2D structure. Most importantly, we propose two evolution mechanisms of thermal transport properties with structure: *in-plane* and *out-of-plane* folding evolution. Note that Janus-graphene produces *out-of-plane* Janus-type buckling superfolding in addition to the *in-plane* 4-6-8 ring folding structure. The *in-plane* and *out-of-plane* double superfolding can suppress the contribution of flexible acoustic (FA) phonons, resulting in low group velocity, weak phonon-phonon scattering, weak phonon hydrodynamic effect, and strong asymmetry, which finally contributes to low $\kappa$. The findings of this investigation have significant ramifications for our comprehension of folded atomic structures' effects on thermal transport, as well as the connection between structure and $\kappa$. The aforementioned findings are anticipated to be crucial in creating and improving the heat transfer capabilities of novel nanomaterials.

## 2. Results and discussion

### 2.1 Evolution mechanism of thermal conductivity with folded structure

To demonstrate the correlation between the folding behavior of crystal structures of different two-dimensional carbon allotropes, Fig. 1 (a) shows two conventional folding evolution paths. All 2D carbon compounds exhibit $sp^2$ bonding, which aims to block the influence of direct $sp^3$ hybrid bonding. From path one, the six membered ring of graphene shrinks to a four membered ring and expands to an eight membered ring, forming the crystal structure of T-graphene. The 4-membered ring in T-graphene further evolved into a Kagome-graphene with 3-membered and 12-membered rings. Similarly, graphene can be folded into Y-biphenylene and biphenylene with 4-6-8 membered rings along the second path. The difference is that the derived phase of biphenylene can fold along the direction of out of plane buckling, forming Janus graphene with non-chemical Janus phase, which has the same 4-6-8 rings as Y-biphenylene and biphenylene.

Intriguingly, the $\kappa$ of folded carbon allotropes is generally lower than that of unfolded carbon allotropes, even one to two orders of magnitude lower than that of pure hexacyclic graphene, as shown in Fig. 1(b). Graphene, a perfect 2D carbon allotrope, exhibits an ultrahigh $\kappa$ of 2970 W/mK *via* machine learning force constant potential (MLFCP) method, serving as an initial research standard due to its perfect planarity and pure six-membered ring framework. The results of MLFCP are similar to our calculated results using DFT (3150 W/mK in Fig. S1(a)) and previous experimental results of~3000W/mK [14,34]. The thermal conductivity of folded T-graphene calculated by DFT method is 645 W/mK, which is lower than that of graphene due to low-lying acoustic phonon molecular folding [18]. Further, Kagome-graphene has a lower $\kappa$ than T-graphene based on DFT method, ~140 W/mK, and is a natural carbon-based phononic crystal with larger pores. Simultaneously, graphene could additionally be folded into 4-6-8 ring biphenylene [33,35], Y-biphenylene, and Janus-graphene. Interestingly, the $\kappa$ of planar biphenylene and Y-biphenylene in the x(y) direction are 267 (421), 330 (500) W/mK, respectively. Due to the nearly similar carbon ring composition, the difference in thermal conductivity between planar biphenylene and Y-biphenylene is not very significant, as has also been reported in previous studies [33,36]. On the contrary, Janus-graphene has a $\kappa$ of 84 W/mK, putting it an order of magnitude lower than biphenylene and Y-biphenylene with the identical 4-6-8 ring.

These results demonstrate that the reduced κ of the same $sp^2$-hybridized 2D carbon material can be attributed to the folding of the in-plane folding of planar crystals, which forms phononic crystal-like carbon allotropes with different atomic chain configurations, thereby reducing κ. The evolution of in-plane folding is similar to that of phononic crystals, where the folding structure leads to the switching of related bandgaps, thereby altering the transport characteristics of phonons. Different ring types lead to different arrangement periodicity and symmetry of phononic crystals, leading to different phonon transport properties and diversifying $\kappa$. Note that for Janus graphene, the lower thermal conductivity compared to biphenylene and Y-biphenylene excludes the influence of in-plane structural folding. Because Janus-graphene, biphenylene, and Y-biphenylene share identical 4-6-8 carbon ring, the difference between the $\kappa$ of biphenylene and Y-biphenylene is negligible. As a consequence, this dramatic difference in Janus-graphene may be traced to its *out-of-plane* superfolding to generate the Janus-configuration, which goes beyond *in-plane* folding to produce stronger phonon-phonon scattering.

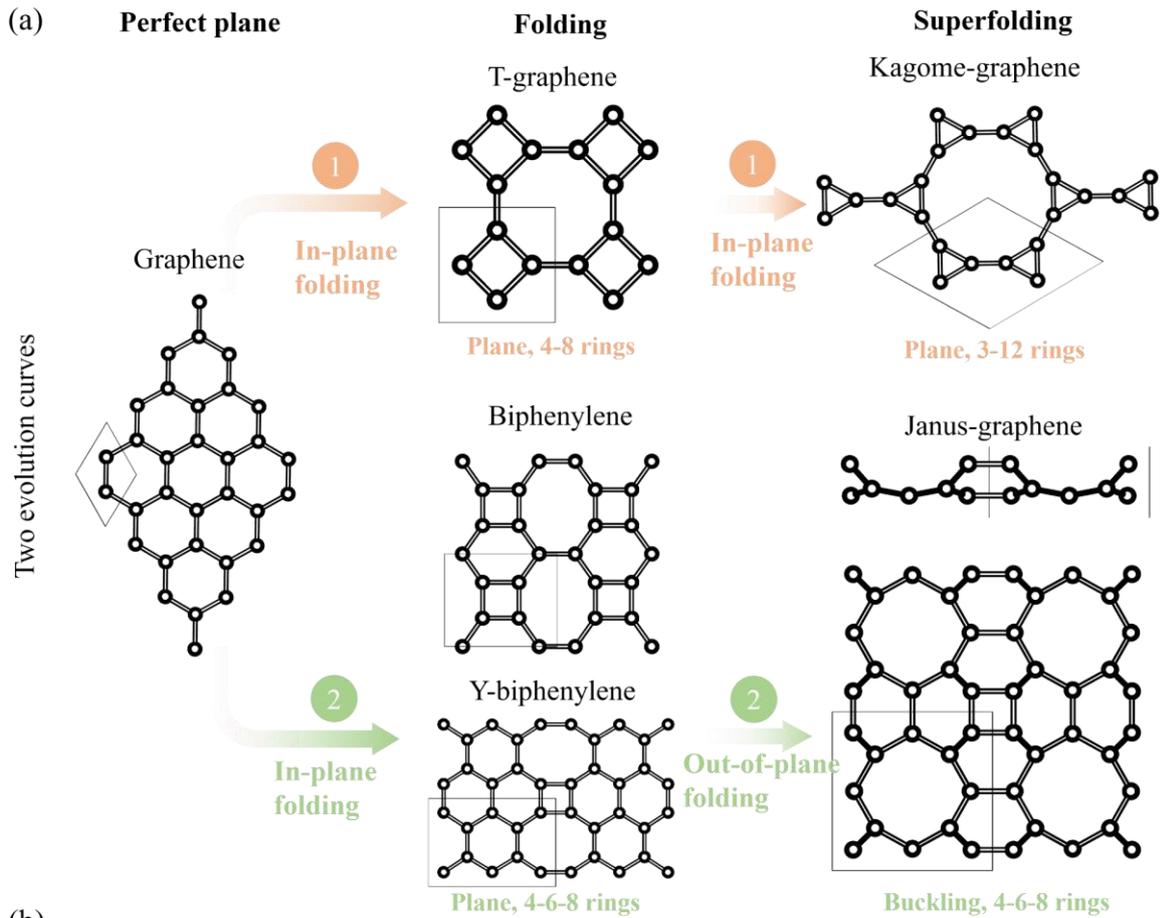

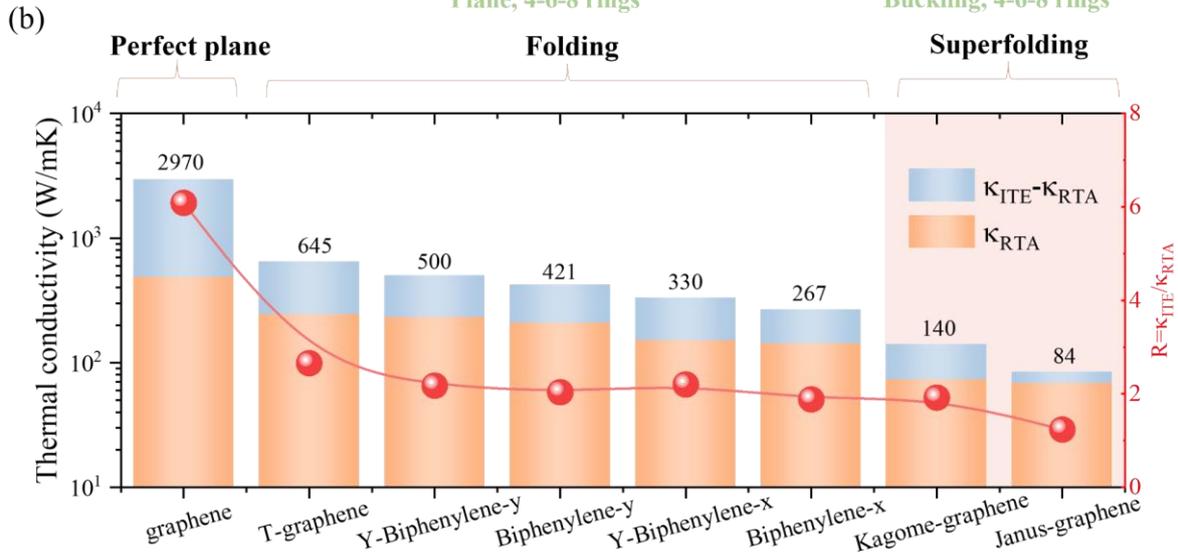

Figure 1. **Evolution trends of structure and thermal conductivity**. (a) Two evolution trends of the structures for 2D carbon crystals with $sp^2$ hybridization along the *in-plane* (orange arrow) and *out-of-plane* (green arrow) directions. (b) Thermal conductivities corresponding to different lattice structures. The thermal conductivity of graphene comes from the machine learning potential model, while the results of other carbon allotropes come from density functional theory calculations. The red line highlights the ratio (R) of the thermal conductivity calculated by the iterative method ($\kappa_{ITE}$) to that calculated by the relaxation time approximation method ($\kappa_{RTA}$).

## 2.2 Machine learning potential combined with phonon Boltzmann transport equation

To explore and understand the underlying fundamental link between the unique configuration of Janus-graphene and phonon transport, machine learning methods were employed to construct interatomic force constant potentials to solve the phonon Boltzmann transport equation (BTE). As shown in the workflow diagram of Fig.2(a), the structures and corresponding energies and forces at different temperatures are calculated by *ab initio* molecular dynamics (AIMD), which are used to construct the data sets required for the simulations. The primary purpose of the interatomic force constant potential, as opposed to the molecular dynamics potential of dynamics, is to compute the low-order and high-order derivatives of energy based on the finite displacement method to solve the BTE, indicating the larger significance for the energy and force of static small displacements. As shown in Figs. 2(b) and (c), the energies and forces predicted by the machine-learned potential (MLP) were compared with the results from first-principles calculations, where the good agreement with each other demonstrated the potential function model's good fitting ability for predicting the energy and force for the perturbed configuration of Janus-graphene. The low root mean square error of energies (RMSE: 18.52 meV) and forces (RMSE 38meV/A) indicates the high quality and strong robustness of the machine learning potential function model. In solving BTE, machine learning networks can map the energies and forces produced by AIMD to obtain low-order and high-order force constants. The third and fourth-order force constants among them are often enough to describe their anharmonicity. As a result, we mainly evaluate the role of three- and four-phonons, especially the former.

In machine learning networks, it is required to concentrate on an atom and encode its domain information, or the descriptor of the local environment. All of these descriptors must satisfy numerous types of symmetry criteria, including translational and rotational invariance, as well as permutation (with regard to the exchange of two atoms of the same sort). Descriptions are usually limited to the local environment of atoms within a specific cut-off radius, typically 5 or 6 Å for covalent bonds, in order to be valid in practice. We reveal the extent of interatomic interactions in the local environment by normalized force constants traces trace (NFCT) in individual neighborhoods. The interatomic force constant can be obtained as the second derivative of the energy $E$ by: [19]

$$\frac{\partial^2 E}{\partial R_i \partial R_j} = \begin{bmatrix} \frac{\partial^2 E}{\partial R_x \partial R_x} & \frac{\partial^2 E}{\partial R_x \partial R_y} & \frac{\partial^2 E}{\partial R_x \partial R_z} \\ \frac{\partial^2 E}{\partial R_y \partial R_x} & \frac{\partial^2 E}{\partial R_y \partial R_y} & \frac{\partial^2 E}{\partial R_y \partial R_z} \\ \frac{\partial^2 E}{\partial R_z \partial R_x} & \frac{\partial^2 E}{\partial R_z \partial R_y} & \frac{\partial^2 E}{\partial R_z \partial R_z} \end{bmatrix}, \qquad (1)$$

$$\text{NFCT} = \frac{\frac{\partial^2 E}{\partial R_{0,x}\partial R_{n,x}} + \frac{\partial^2 E}{\partial R_{0,y}\partial R_{n,y}} + \frac{\partial^2 E}{\partial R_{0,z}\partial R_{n,z}}}{\frac{\partial^2 E}{\partial R_{0,x}\partial R_{0,x}} + \frac{\partial^2 E}{\partial R_{0,y}\partial R_{0,y}} + \frac{\partial^2 E}{\partial R_{0,z}\partial R_{0,z}}}, \tag{2}$$

where $\frac{\partial^2 E}{\partial R_{i,x}\partial R_{j,x}}$ means atomic interaction forces along the *x*-direction between the $i_{\text{th}}$ and $j_{\text{th}}$ atoms and labeled "0" means the specified initial atom. As shown in Fig. 2(d), the trained machine-learned potential can well reflect the atomic interactions originating from the second derivative of the potential well. It additionally demonstrates that distance-dependent harmonic and anharmonic interactions can arise at the cutoff radius of 6 Å.

To validate the accuracy of the machine learning potential function model, we firstly construct the phonon dispersion curve of Janus-graphene and compare to density functional theory (DFT) results. The influence of different supercells on phonon dispersion was tested in Janus-graphene, and the 4×4×1 supercell was finally selected as shown in supplementary Fig. S2. The results obtained by the MLP model are in good agreement with those computed using DFT, as shown in Fig. 2(e, f). The distinction is that Janus-graphene has more phonon branches than graphene, resulting in additional phonon modes, which is due to the asymmetry in the inherent structure of Janus-graphene. The acoustic phonon branch is significantly suppressed, demonstrating strong phonon softening. As a result, this leads to a significantly weakened contribution of the acoustic phonon branch that dominates the $\kappa$, contributing to an ultralow $\kappa$, as explained later. Meanwhile, as shown in Fig. 2(g), the three acoustic phonon branches in Janus-graphene are suppressed below 10 THz, where low-frequency phonons usually dominate. Acoustic phonons are frequency-confined, resulting in lower $\kappa$ than graphene, possibly related to the asymmetry induced by *out-of-plane* folding. This asymmetry impedes or suppresses acoustic phonon mode transmission due to the Janus configuration. The convergence of multiple parameters, such as supercell size, grid, and anharmonic cut-off radius, is presented in Fig. 2(h) to test the stability of the $\kappa$ computed by the potential function. Under diverse settings, the average $\kappa$ of Janus-graphene converges to 87±13 W/mK, indicating a reliable convergence result. The above results confirm that machine learning combined with the force constant potential can accurately solve the phonon BTE to obtain reasonable phonon transport properties for Janus-graphene, which ensures the validity of further in-depth physical information.

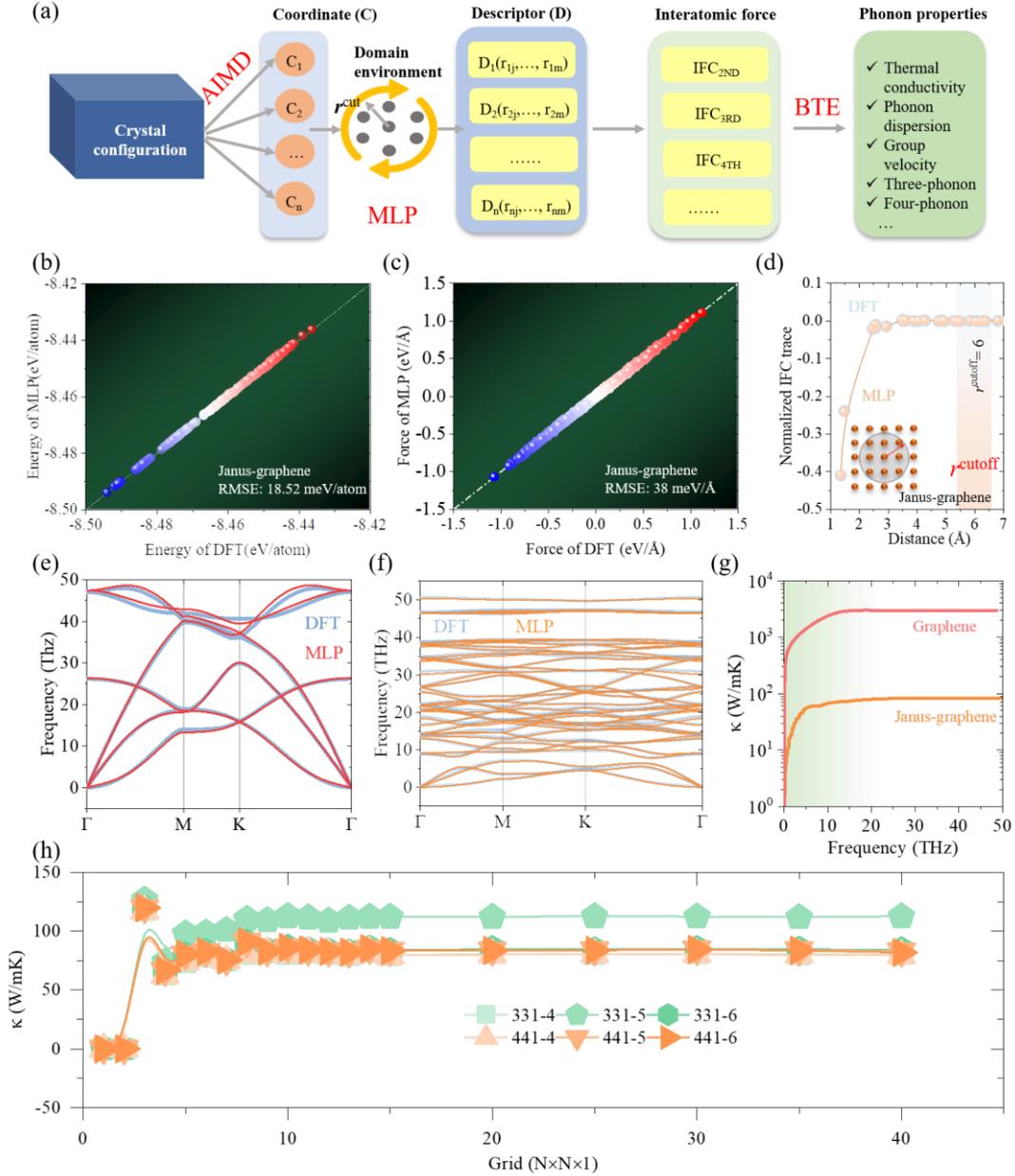

Figure 2. **Machine learning force constant potentials and density functional theory (DFT) calculations.** (a) Workflows for thermal transport properties prediction. (b) Energy, (c) force and (d) normalized force constant traces of Janus-graphene. Phonon dispersion in (e) graphene and (f) Janus-graphene. (g) Cumulative thermal conductivity of graphene and Janus-graphene. (h) Thermal conductivity convergence test for Janus-graphene. Taking "331-4 (or 441-6)" as an example, it means the thermal conductivity result is calculated using a 3×3×1 (or 4×4×1) supercell and the 4$^{th}$ (6$^{th}$) nearest neighbor. Thermal conductivity exhibits good convergence when using a $4 \times 4 \times 1$ supercell under the 6th nearest neighbor, passing through a $40 \times 40 \times 1$ grid.

## 2.3 Three-phonon and Four-phonon effects

To further reveal the origin of $\kappa$ with structural evolution, we will comprehensively compare the evolution of $\kappa$ and phonon transport behavior, using Janus-graphene and graphene as typical cases. Mode-level $\kappa$ and relaxation time as functions of Brillouin zone position and frequency are depicted in Figs. 3(a) and (b). The phonons that contribute significantly to $\kappa$ reside mostly in the low-frequency area below 10 THz, namely the flexible (FA) and transverse acoustic (TA) branches. As shown in Fig. 3(c), the temperature-dependent behavior of graphene and Janus-graphene was determined using the Iterative (ITE) and relaxation time approximation (RTA) methods. Based on the ITE approach, the $\kappa$ at 300K for Janus-graphene and graphene are 84 and 2790 W/mK, respectively, which are in good agreement with the DFT results as shown in Supplementary Fig. S1. In addition, we also compare the robustness of the force constants for graphene as shown in Supplementary Note S1, where the robust force constants from the MLP agree well with the DFT results. These results demonstrate the accuracy and robustness of machine learning-driven force constant potentials. Note that the $\kappa$ of Janus-graphene is almost two orders of magnitude lower than that of graphene due to the double symmetry breaking caused by *in-plane* and *out-of-plane* superfolding. However, only three-phonon scattering is considered here, and four-phonon scattering in Janus-graphene should also be considered.

Building on the study of Mortazavi [37] and Han et al. [38], we here reformulate the four-phonon interface (*Fourthorder_MTP.py*) between the machine learning potential and the phonon BTE solver. The influence of the four-phonon effect on graphene $\kappa$ under DFT and MLP is compared to reveal the accuracy of the MLP-Fourphonon process. As shown in Supplementary Fig. S3, the two can be well-matched and are consistent with the previous results of Feng et al. [16], revealing a high-precision description of higher-order force constants by the machine learning model. In addition, the mode-level group velocity and relaxation time considering three-phonon and four-phonon mechanisms are compared in Supplementary Fig. S4, where the results of the MLP method and the DFT method are in good agreement. The impact of four-phonon scattering on $\kappa$ has been deeply discussed in some studies [14,16,34]. The role of four phonons may be significant at high temperatures or in some special cases, such as flat bands in $AgCrSe_2$ [39] or huge acoustic-optical band gaps in BAs [40–44]. Han and Ruan et al. found that after considering four phonons, the thermal conductivity of graphene is approximately ~1300 W/mK [45], which is slightly higher than the previous results of Feng et al. [16] and our MLP method (~900W/mK). The slightly higher reason may be due to the slightly sparse grid grid, so we provide a denser grid-test for thermal conductivity considering four-phonon scattering in Fig. S5 in the supplementary material. However, the $\kappa$ after considering the four-phonon scattering in Janus-graphene is 52 W/mK, but this is not a substantial order of magnitude reduction, indicating that the three-phonon scattering still plays a dominant role in the phonon-phonon scattering process. $P_3$ and $P_4$ represent the scattering phase spaces of three and four phonons as a dimensionless parameter, respectively, which measure the scattering probability of three and four phonons, respectively. As shown in Fig. 3(d), $P_3$ is significantly higher than $P_4$ in Janus-graphene, especially at low frequencies, revealing the origin of relatively weak four-phonon scattering. Hence, the $\kappa$ calculated based on the three-phonon effect is still acceptable in the subsequent analysis.

The contribution of different phonon branches for Janus-graphene to the $\kappa$ is shown in Fig. 3(e). Since phonons follow the Bose-Einstein distribution, the contributions of all phonon branches decrease with increasing temperature in the temperature range of 200-800 K. The $\kappa$ of Janus-graphene at 300 K is mostly provided by the FA, TA, and optical branches. The TA branch has contributed most, reaching 36%, which is substantially higher than graphene. At 300K, the contribution percentage of FA branch in graphene approaches 80%, nearly fully governing $\kappa$, but the FA branch in Janus-graphene occupies just 26%, a considerably fewer proportion than in graphene. Although after considering four phonons, the contribution of FA phonons in graphene can still be as high as 75%. [45] This indicates that the Janus configuration significantly weakens the contribution of the FA branch, leading to a decrease in $\kappa$. The contribution of the optical phonon branch to $\kappa$ in Janus-graphene provides 27% at 300 K, which is only slightly lower than that of the TA branch. In contrast, the contribution of optical branches to $\kappa$ in graphene is almost negligible. Since the presence of more phonon branches than graphene depresses the contribution of the FA branch, the optical branches of Janus-graphene show a significantly enhanced contribution to $\kappa$, especially at high temperatures.

Furthermore, the coupling between the mode-level phonon properties in the frequency domain interval is plotted in Figs. 3(d-f) for graphene and Janus-graphene, which mainly highlights the contribution distribution of different phonon properties to the frequency-dependent $\kappa$. In graphene, the FA branch has a particularly high $\kappa$ density compared to other branches, which is mainly due to its weak phonon-phonon scattering rather than group velocity. The group velocity of the FA branch with quadratic dispersion is almost zero in the center of the Brillouin zone, which has led to the long-standing misperception that graphene has a relatively low $\kappa$ because zero group velocity leads to low $\kappa$ in FA branch. But the high relaxation time of graphene can significantly enhance the thermal conductivity of FA branch, regardless of whether four phonons are considered, as shown in Figs. 3(e) and (f). However, In Janus-graphene, the $\kappa$ density of the FA branch significantly decreases while the $\kappa$ density of the TA branch increases significantly. The reduction of the FA branch is derived from both the enhanced phonon-phonon scattering contributed by the FA branch and its lower group velocity. The TA branch of Janus-graphene has a stronger group velocity than the FA branch, which can significantly increase the contribution of $\kappa$ because the mode-level $\kappa$ is proportional to the square of the group velocity. Overall, the Janus configuration weakens the contribution of the FA branch to $\kappa$ in Janus-graphene, making it not the sole dominant thermal transport property.

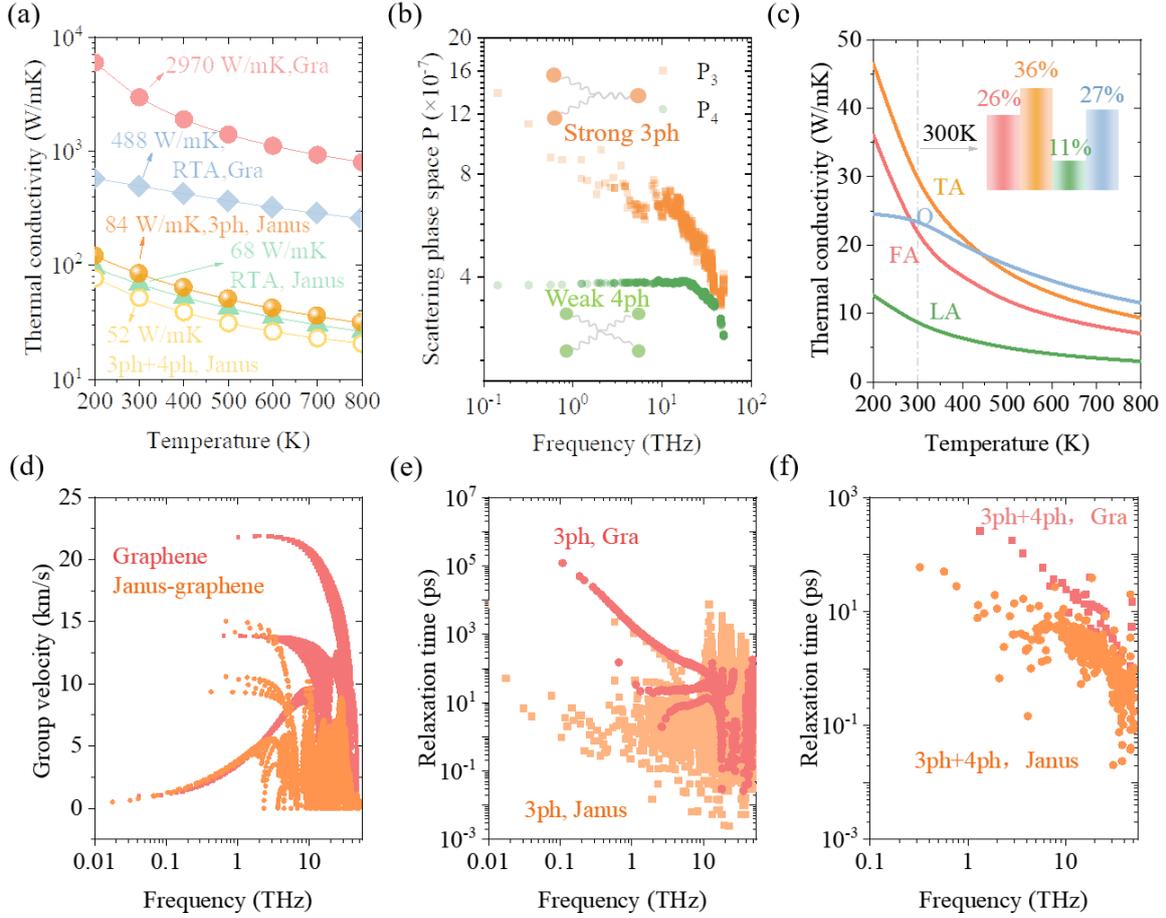

**Figure 3. Thermal transport properties.** (a) Temperature dependent thermal conductivity in graphene and Janus-graphene. (b) Three- and four-phonon scattering in phase space. (c) Temperature-dependent thermal conductivity contribution for different branches. (d) Brillouin zone dependence of group velocities for machine learning potential methods and density functional theory methods. (e) Mode-level relaxation times under three-phonon scattering for machine learning potential methods and density functional theory methods. (f) Mode-level relaxation times under three-phonon and four-phonon scattering for machine learning potential methods and density functional theory methods.

**2.4 Phonon properties and phonon-phonon scattering**

Further, representative phonon properties and scattering channels of FA, and TA branches for graphene and Janus-graphene are visualized in the Brillouin zone as shown in Fig. 4, which truly reflects the actual phonon-phonon scattering. The distributions of phonon group velocities for the FA branch in graphene and the FA and TA branches in Janus-graphene are plotted in Figs. 4(a), (b), and (c). Obviously, the phonon group velocity of the FA branch in Janus-graphene is much lower than that of graphene, contributing to the lower $\kappa$. In addition, the TA branch of Janus-graphene has higher phonon group velocities than the FA branch due to the linear dispersion relation, indicating that the TA branch contributes relatively more to the $\kappa$. Meanwhile,

Figs. 4(d), (e), and (f) reveal the phonon scattering rate distributions of FA branch in graphene and FA and TA branches in Janus-graphene. The phonon scattering rates of FA and TA branches in Janus-graphene are almost an order of magnitude, but both are significantly higher than those of FA branch in graphene. The variation of phonon group velocity and scattering rate leads to the weakening of the contribution of the acoustic FA branch in Janus-graphene.

The weakening of the FA branch is structurally due to the reorganization of the phonon dispersion caused by the folding of the *in-plane* structure, which was observed in previous studies [18,31]. But it also notes that *out-of-plane* buckling superfolding also suppresses the FA branch as shown in Supplementary Fig. S6, where the frequency reduction of the acoustic phonon branch exhibits the phenomenon of phonon softening, *i.e.*, the strong anharmonicity is due to the superfolded structure. Further insight into the effect of spatial superfolding on anharmonicity can be obtained through the phonon-phonon scattering channel as shown in Figs. 4(g), (h), and (i). In the three-phonon process, the anharmonic third-order force constant can be obtained with $n = 3$:

$$E_{\alpha_1 \ldots \alpha_n}(i_1 j_1; \ldots; i_n j_n) = \frac{\partial^n E}{\partial u_1(i_1 j_1), \ldots, \partial u_n(i_n j_n)}\bigg|_0, \tag{3}$$

where E is lattice potential energy, and $\vec{u}_{ij}$ represents the perturbation displacement of the *j*th atoms in the *i*th unit. The third-order force constant can be mapped to itself in the *in-plane* when $n=3$ for 2D materials: [19]

$$E_{\alpha_1, \ldots, \alpha_n}(i_1 j_1; \ldots; i_n j_n)(-1)^m = E_{\alpha_1, \ldots, \alpha_n}(i_1 j_1; \ldots; i_n j_n), \tag{4}$$

where *m* comes from the symmetry operation and depends on the scattering phase space. It satisfies:

$$E_{\alpha_1, \ldots, \alpha_n}(i_1 j_1; \ldots; i_n j_n) = 0, m \ odd \tag{5}$$

$$E_{\alpha_1, \ldots, \alpha_n}(i_1 j_1; \ldots; i_n j_n) \neq 0, m \ even \tag{6}$$

The phonon-phonon scattering of the FA branch in graphene [Fig. 4(g)] only involves the even-numbered FA phonons, *i.e.,* FA+FA→TA/LA. The FA branch scattering channels involving odd numbers are suppressed due to the strong planar symmetry driven by the planar structure, which makes the force constant $E_{\alpha_1, \ldots, \alpha_n}(i_1 j_1; \ldots; i_n j_n)$ vanish when mapped to itself (*m = odd*). Differently, the odd number of FA mode scattering channels of FA+FA→FA and FA→FA+FA are allowed in Janus-graphene as shown in Fig. 4(h) owing to the broken selection rule by Janus-structure (*m = even*). In addition, the scattering rate contributed by the phonons of the TA branch is almost the same as that of the FA branch, where the two TA phonons and FA phonons participate in the phonon-phonon scattering as shown in Fig. 3(g). This also explains the same contribution of TA and FA branches to the *κ* [Fig. 3(e)]. Note that the strong anharmonicity in Janus-graphene can be attributed to the fact that the strong asymmetry violates the selection rule of phonon-phonon scattering. Differently, although the *in-plane* symmetry of biphenylene and Y-biphenylene is not protected, the *out-of-plane* symmetry is preserved, leading to lower *κ* than graphene but higher than Janus-graphene [Fig. 1(b)].

The *out-of-plane* folding effect in Janus-graphene leads to a significant breaking of the central and mirror symmetries, inducing a stronger anharmonicity.

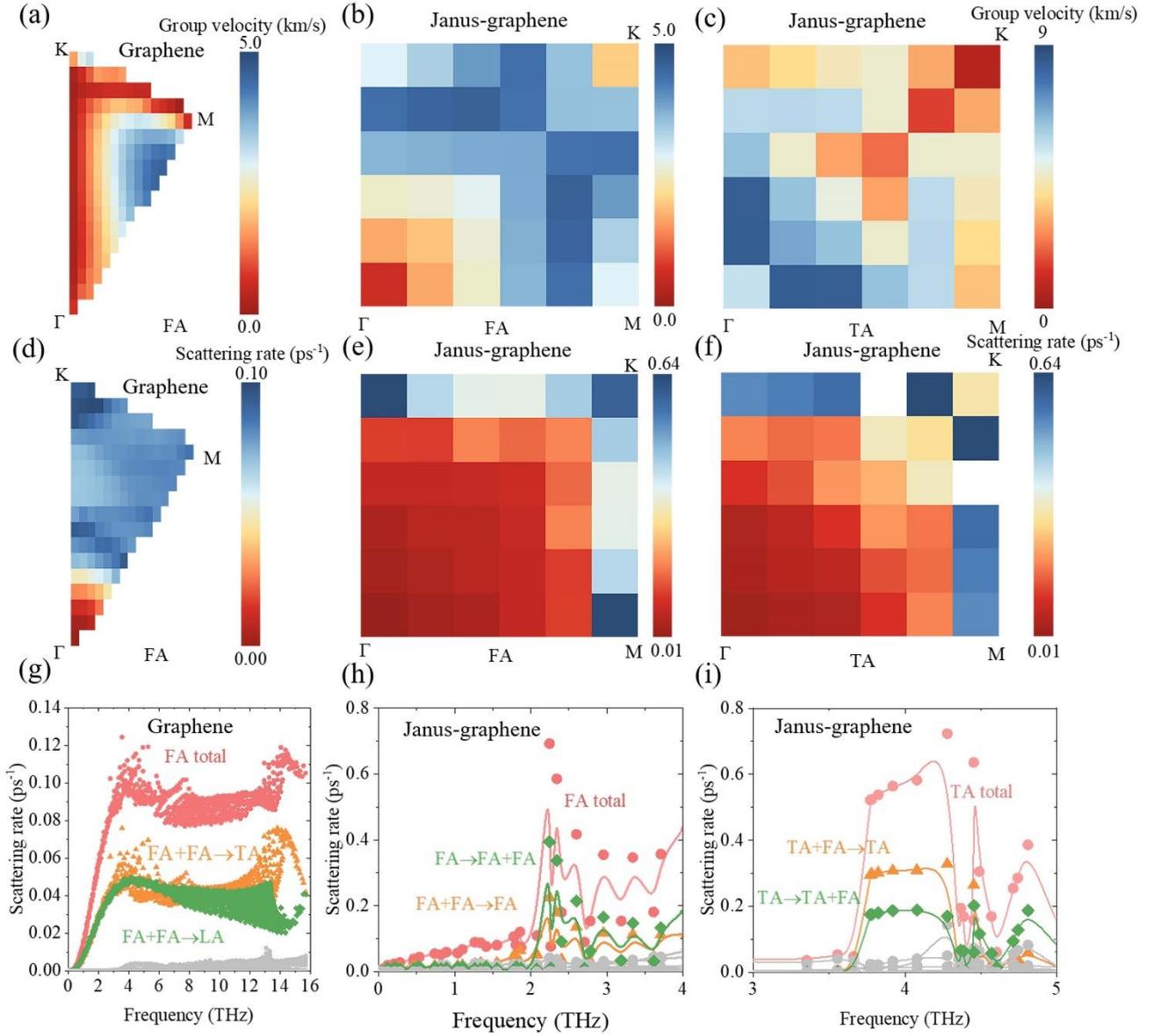

Figure 4. **Phonon properties and phonon-phonon scattering.** Phonon group velocity along highly symmetric paths in (a) graphene and (b, c) Janus-graphene. Phonon-phonon scattering rates along highly symmetric paths in (d) graphene and (e, f) Janus-graphene. Mode-level phonon-phonon scattering channels in (g) graphene and (h, i) Janus-graphene. "FA", "TA", and "LA" represent flexible, transverse and longitudinal acoustic phonon branches.

## 2.5 Phonon hydrodynamic effects

As we know, phonon behavior is largely constrained by the crystal structure of the material itself. As a result of their softness and bending stiffness, 2D materials are easily deformed in the plane, giving rise to flexural acoustic modes with quadratic power, *i.e.*, FA branch. The generation of the FA branch mainly involves the bending vibration of atoms inside 2D materials, similar to the bending modes of thin films. Due to the perfectly planar structure of graphene, these phonons have very little scattering in the direction perpendicular to its plane, *i.e.*, they can propagate freely within the plane. This leads to a high density of states for long-wavelength FA phonons, i.e., a kinetically abundant presence of this type of phonons. In graphene, the collective motion of these long-wavelength FA phonons can exhibit a dynamic flow similar to fluid drift, which is known as the phonon hydrodynamic effect [46–50]. Attenuation of the prominent FA branches in Janus-graphene also predictably reduces this phonon hydrodynamic effect. The difference in phonon hydrodynamics in graphene and Janus-graphene can be intuitively compared by ITE and RTA methods, as shown in Fig. 1(b). The strength of phonon hydrodynamics can roughly be assessed by comparing the $\kappa$ of the ITE and RTA methods ($H=\kappa_{ITE}/\kappa_{RTA}$) [19]. The smaller the $H$ value, the weaker the hydrodynamic effect. As shown in Fig. 1(b), the ultralow H value in Janus-graphene reveals its weak hydrodynamic effect.

The contribution of the relaxation times of graphene and Janus-graphene under the ITE and RTA methods can further reveal a significant difference in $\kappa$, as shown in Fig. 5(a), where the relative magnitudes of the phonon relaxation time (RT) in the ITE and RTA methods can be measured by their ratio ($RT_{ITE/RTA}$). The $RT_{ITE/RTA}$ value of graphene is significantly higher than that of Janus-graphene and higher than 1, revealing the enhanced effect of phonon hydrodynamics on the thermal transport properties of graphene, while it has little effect on Janus-graphene. The significant difference in $\kappa$ can be attributed to the significant weakening of the contribution of the *out-of-plane* acoustic phonon branch FA, and the significant weakening of the FA branch by the iterative method can also be captured in Janus-graphene due to the superfolded structure [Fig. 3(e)]. The significant phonon softening of the FA branch in Janus-graphene compared to graphene can also be trapped as shown in Figs. 2(b) and (c).

The suppression of phonon hydrodynamics in the Janus-graphene can be further attributed to the momentum redistribution during phonon-phonon scattering. Based on the conservation of momentum, the phonon scattering process can be divided into two categories: normal (N) scattering and umklapp (U) scattering:

$$\vec{q} \pm \vec{q'} = \vec{q''} + \vec{K}, \qquad (7)$$

where $\vec{K} = 0$ and $\vec{K} \neq 0$ correspond to N and U processes, respectively. In N-scattering, a collision occurs between two phonons, but the sum of their wave vectors remains within the unit vector of the reciprocal lattice. This scattering event does not change the wave vector magnitude and direction of the phonons, only their phase, leading to the fact that N scattering events preserve momentum. However, in U-scattering, the scattering between two phonons produces a wavevector that exceeds the unit vector of the reciprocal lattice. This means that the sum of the wave vectors has crossed the boundary of the first Brillouin zone and cannot be represented by the primitive cell vectors of the lattice. In this case, the excess momentum cannot be retained between the phonons, but is transferred to other parts in the lattice. Thus, the U-scattering event results in a loss of momentum. In phonon hydrodynamics, the relative contributions between N and U processes determine the propagation and depletion behavior of phonons, whose relative strength can be measured by their ratio (N/U). As shown in Fig. 5(b), the low-frequency phonons in Janus graphene contribute to the weak N process, leading to a weak phonon hydrodynamic effect due to its significantly suppressed FA branch. Obviously, compared with graphene, Janus-graphene exhibits a weaker phonon hydrodynamic effect, where its N- and U-process phonon scattering rates are almost the same (the ratio is approximately equal to 1), which suggests that the superfolding of the atomic structure can significantly suppress the phonon hydrodynamic effects in Janus-graphene.

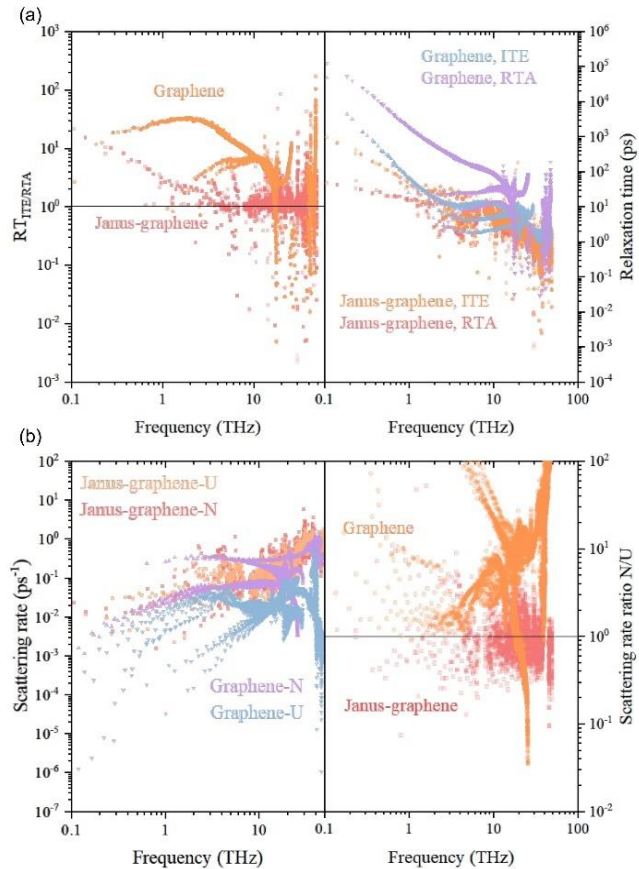

Figure 5. **Phonon hydrodynamic effect**. (a) Phonon relaxation times comparison between iterative and relaxation time approximation methods. (b) Phonon-phonon scattering for umklapp (U) and normal (N) processes determined by conservation of momentum.

**2.6 Electronic properties and anharmonicity**

The superfolded buckled structure significantly suppresses the contribution of FA branch to thermal conductivity, and this structure-function evolution enables us to gain deep insights from atomic interactions and anharmonicity. In graphene, two carbon atoms form a strong and stable covalent bond in pure $sp^2$ hybridization by sharing electron pairs. In $sp^2$ hybridization, one 2$s$ orbital and two 2$p$ ($p_x$ and $p_y$) orbitals of carbon atoms are linearly combined to form three equivalent $sp^2$ hybridization orbitals. These $sp^2$ hybrid orbitals form three σ bonds in graphene, which lie in a plane and are separated by an angle of 120° from each other. The remaining one $p_z$ orbital is perpendicular to the plane and forms a π bond, responsible for the π electron conjugated system in graphene. This ideal electronic configuration exhibits strong symmetry, which limits the anharmonicity in the atomic vibration process due to the perfectly planar structure. Differently, the *out-of-plane* buckling significantly destroys the planar features, especially the $p_z$ orbitals of Dirac cones (for graphene), making the $p_z$ electrons localized in the *out-of-plane* direction to form a quasi-$sp^3$ hybridization as shown in Fig. 6(a), *i.e.*, dangling bond [51,52]. A dangling bond refers to a covalent bond around an atom in a material that is not fully saturated, resulting in the presence of unpaired electrons on the atom, which makes the atom more reactive. This usually happens with surface or edge atoms because they lack neighbors adjacent to interior atoms. In order to stabilize a surface or interface, it is usually necessary to saturate the dangling bonds by different methods, such as adding surface modifiers [53], hydrotreating [54], or using other chemicals [55] to fill these unsaturated bonds. Such modification can also significantly suppress its anharmonicity and enhance thermal transport properties [56,57].

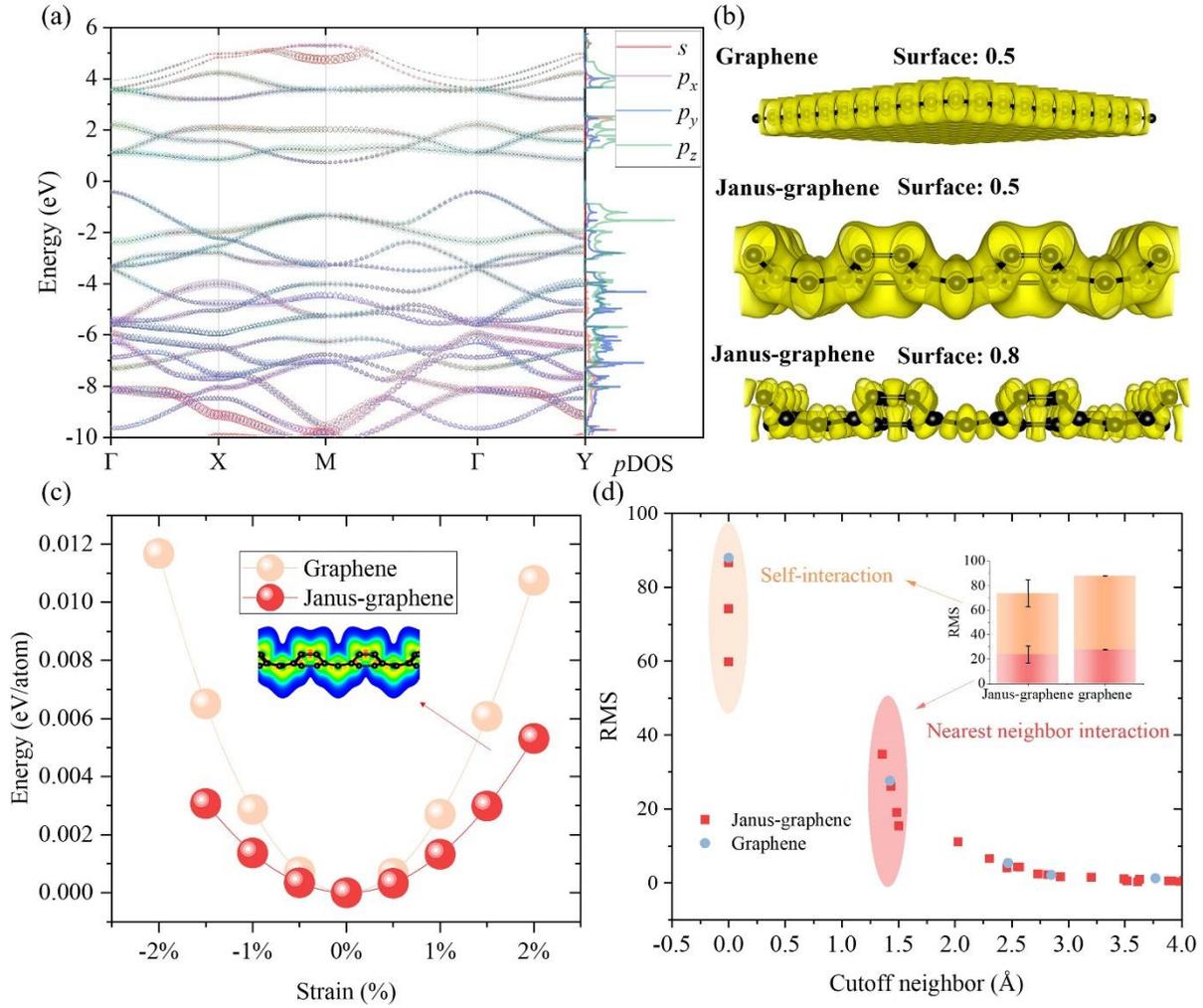

Figure 6. **Electronic properties and lattice anharmonicity**. (a) Orbital hybridization of Janus-graphene. (b) 3D dimensionless electron localization function comparison of graphene and Janus-graphene. (c) Comparison of atomic vibration potential wells of graphene and Janus-graphene. (d) Comparison of atomic vibration RMS displacement of graphene and Janus-graphene. Embedded are the averaged self-interactions and nearest-neighbor interactions.

However, limited by the 2D features of Janus-graphene and its unique superfolded configuration, the prominent dangling bond states in Janus-graphene are activated, demonstrating strong stereochemical activity. The diverse electron localization behavior can be further revealed by the electron localization function (ELF): [51]

$$\text{ELF} = \left(1 + \left\{\frac{K(r)}{K_h[\rho(r)]}\right\}^2\right)^{-1}, \tag{8}$$

where where $K_h[\rho(r)]$ is also the normalization coefficient of $K(r)$ and is the curvature in a homogeneous ($h$) electron gas of density $\rho$ in position $r$. The size of the ELF represents the distribution of electrons from

**18** / **28**

fully delocalized (ELF=0) to fully localized (ELF=1). When the ELF value is greater than 0.5, it means that the electrons are more localized in a certain position. As shown in Fig. 6(b), the ELF potential energy surfaces of graphene and Janus-graphene exhibit different localized morphologies due to the unique hybridization evolution. The localized electrons in Janus-graphene hang in the *out-of-plane* direction of carbon atoms, which is also revealed by the inset of the 2D ELF in Fig. 6(c). In thermal vibrations, the homogeneously distributed ELF in graphene lead to weak anharmonic interactions in wavefunction overlap due to strong symmetries. In contrast, the pronounced asymmetry in the Janus-graphene induces strong nonlinear Coulomb forces due to *out-of-plane* buckling superfolding, leading to its strong anharmonicity. Strong anharmonicity enables further insight to be gained through the anharmonic potential well shown in Fig. 6(c), where Janus-graphene exhibits a wider potential energy surface at the same amplitude, indicating that atoms need more energy to recover during vibration. Strong asymmetry and anharmonic potential wells can be reflected in the interatomic force constants to affect thermal transport properties, which can be quantified by the root mean square of atomic interactions: [59]

$$RMS = \left[\frac{1}{9}\sum\left(\frac{\partial^2 E}{\partial R_{i,\alpha}\partial R_{j,\beta}}\right)^2\right]^{\frac{1}{2}}, \quad (9)$$

where RMS can be obtained by traversing each atom with the initial central atom, representing the interatomic interactions. As shown in Fig. 6(d), the RMS distribution of Janus-graphene is more inhomogeneous compared to that of graphene, indicating non-uniform atomic repulsion in wave function overlap. This inhomogeneity of RMS originates from the inhomogeneity of the force constant, leading to the renormalized phonon-phonon scattering, which finally leads to a significant decrease in the $\kappa$ of Janus-graphene.

**2.7 Universal insights into asymmetric interactions**

Ultimately, it could potentially be feasible to arrive at broad conclusions about the causes of the uneven distribution of force constants from the folding evolution of atomic geometries. As shown in Figs. 7(a) and (b), carbon atoms are organized in a perfect hexagon to create planar graphene, and the bond angles between carbon atoms determine how the material is structured. In graphene, each carbon atom makes a covalent link with its three neighbors by pure *sp*$^2$ hybridization, leading to bond angles that are precisely 120°. This optimal hexagonal arrangement causes graphene to have a honeycomb lattice structure on the plane, with carbon atoms grouped in a close-packed way, enabling a highly symmetrical structure. As a result, the nonlinear Coulomb repulsion force (or non-harmonic interaction) of the bonding and non-bonding electrons in graphene is easily canceled or annihilated during the thermal vibration of atoms. In other words, *out-of-plane* phonon-phonon scattering is governed by a symmetry in which the odd number of FA phonons involved are mapped onto themselves and prohibited by the phonon selection rule. However, when the pure 6-membered ring is transformed into a 4-6-8 ring, the bond angle deviates from the ideal 120° in the *in-plane* direction. For biphenylene and Y-biphenylene [Fig. 1(a)], the *in-plane* bond angles are approximately 90°, 120°, and 180°, leading to the breaking of the central rotational and translational symmetry and thus the redistribution of the Coulomb repulsion between the in-plane charges. Among them, phonon-phonon scattering channels previously suppressed by symmetry are activated, especially those involving FA phonons. As shown in Figs. 7(c) and (d), when the *out-of-plane* configuration is introduced, the Coulomb repulsion along the *out-of-plane* direction is renormalized, resulting in a further broken symmetry. As a result, the role of the FA branch in Janus-graphene is further weakened, and no longer dominates the $\kappa$, while FA and TA co-dominate. Combined with suppressed phonon group velocity, strong phonon scattering and weak phonon hydrodynamic effects lead to lower $\kappa$. Therefore, the $\kappa$ of Janus-graphene is not only lower than that of graphene, but also about an order of magnitude lower than that of biphenylene and Y-biphenylene [Fig. 1(b)]. Because the bond length variation is not as substantial owing to the identical carbon atoms, the geometric bond angle distribution has a major impact on the strong asymmetry, leading to various types of in-plane and out-of-plane Coulomb repulsion distributions, as illustrated in Fig. 7. Such a general insight can be widely used to explain the thermal transport properties in other 2D carbon allotropes, such as D-graphene, T-graphene and Kagome-graphene [18,31,60]. The potential mapping link between atomic-scale geometry and phonon level can also serve as a design foundation for thermally functional materials, as well as valuable clues for the prediction and regulation of thermal transport properties of abundant 2D carbon allotropes or carbon-based phononic crystals.

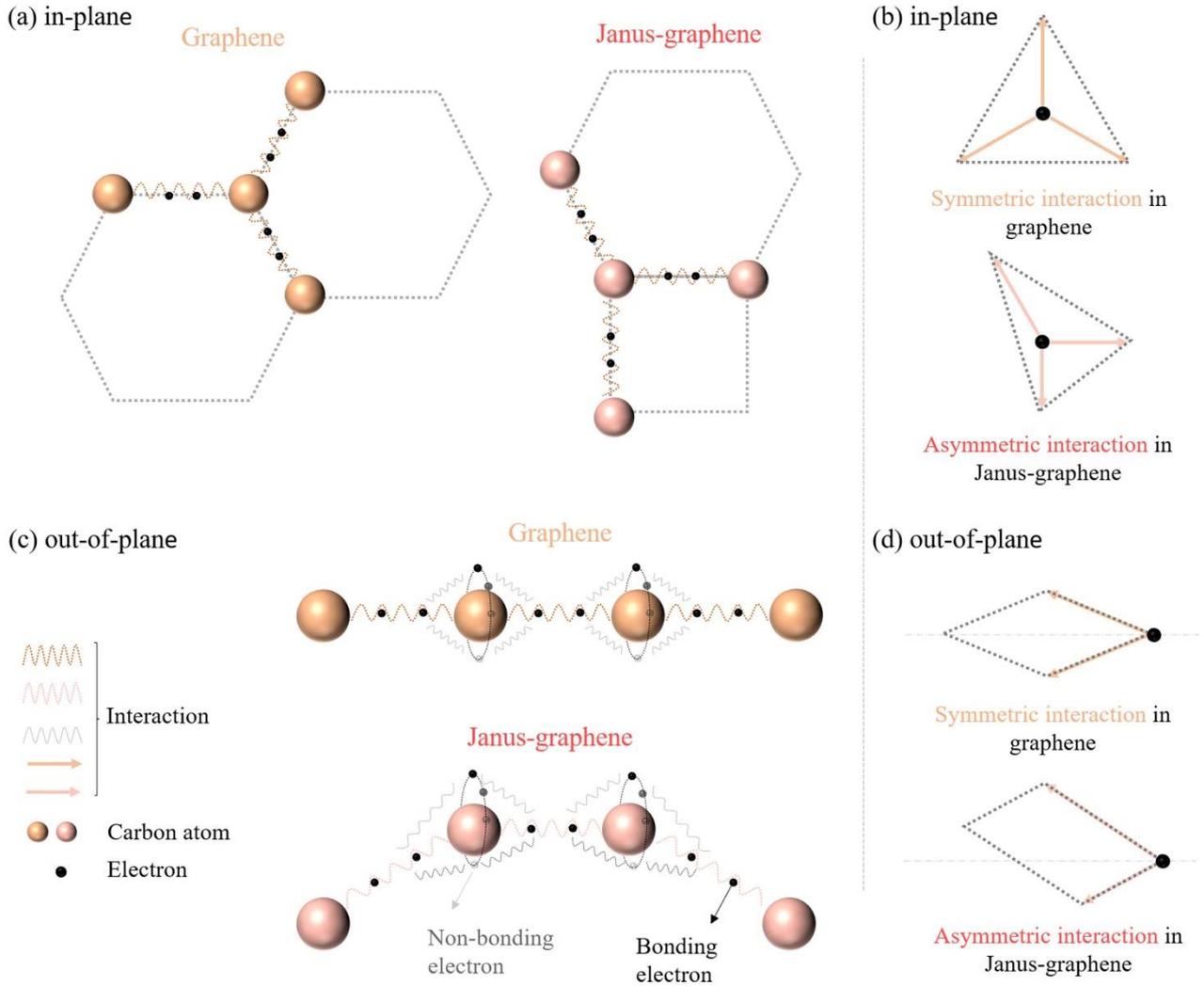

Figure 7. **Folded structure-induced asymmetric atomic interactions.** Schematic diagram of (a) atomic structure and (b) atomic interactions along the *in-plane* direction. Schematic diagram of (a) atomic structure and (b) atomic interactions along the *out-of-plane* direction.

## 3. Conclusion

In summary, we take advantage of the powerful fitting ability of machine learning methods to solve the phonon BTE by establishing a regression potential energy surface from DFT to MLP, thereby obtaining the thermal transport properties of Janus-graphene and graphene. We provide an interface module between MLIP and ShengBTE with respect to the fourth-order force constant, which further isolates and accounts for four-phonon scattering effects compared to previous molecular dynamics potentials [30]. Moreover, the molecular dynamics method cannot consider the quantum response, but the machine learning force constant potential method can further consider the quantum effect, *i.e.*, the phonon-phonon scattering of various orders. In addition, due to the limitations of the finite displacement method, asymmetric systems often consume massive computing resources to calculate small displacement supercells, which increase exponentially with the order of the force constant. The powerful fitting ability of machine learning methods can quickly draw accurate potential energy surfaces for different configurations through the optimization of data sets [61], so as to quickly obtain higher-order force constants. Thus, the machine learning force constant potential method can provide a more convenient method for exploring the high-order phonon-phonon scattering effects in complex asymmetric systems.

More importantly, we propose two evolution paths of folded structure and thermal transport properties for pure $sp^2$ hybridized 2D carbon structures, namely *in-plane* and *out-of-plane* folds. The $\kappa$ of T-graphene and Janus-graphene, which consist of superfolded evolved structures, are significantly lower than those of other 2D carbon allotropes with pure $sp^2$ hybridization. Among them, T-graphene undergoes superfolding of structural evolution in the plane, thereby forming a natural phonon crystal-like carbon configuration with 3-12 carbon rings. Differently, Janus-graphene produces *out-of-plane* Janus-type buckling superfolding in addition to the *in-plane* 4-6-8 ring folding structure, leading to a $\kappa$ value one order of magnitude lower than that of biphenylene (and Y-biphenylene) and two orders of magnitude lower than that of planar graphene. A detailed physical picture can be obtained through Janus-graphene's structural evolution, mode-level phonon properties, momentum-conserving phonon-phonon scattering, electronic structure, and atomic interactions with graphene as a benchmark. The superfolding of the Janus structure suppresses the contribution of the FA phonon branch, leading to low group velocity, strong anharmonicity, weak phonon hydrodynamic effect, and strong asymmetry, which ultimately contributes to low $\kappa$. This study sheds new light on understanding the relationship between the structure and function of 2D materials, especially heat transport, and for designing materials with specific thermal transport properties.

## 4. Method

The initial data set was constructed using the Vienna ab-initio simulation package (VASP) [62] based on density functional theory as the input of the machine learning network. In *ab initio* molecular dynamics, the 4×4×1 supercell is constructed to avoid the influence of periodic boundary conditions on the system. When the kinetic energy cutoff is1000 eV, a 2×2×1 Monkhorst-Pack [63] k-mesh is used to sample the Brillouin zone until the energy and Hellmann-Feynman force accuracy converge to $10^{-5}$ eV and $10^{-4}$ eV/Å, respectively. Based on the isothermal and isobaric (NPT) ensemble, 1000 frames of data are extracted in 50K, 150K, 300K, 500K, and 700K, respectively, when the time step is 0.1 fs.

To fit the force constants, we used a linear programming-based Machine-Learning Interatomic Potential (MLIP) developed by Shapeev et al. [64] to develop interatomic potential functions. The total energy of the configuration ($E_m^{MLP}$) is given by a linear combination of a set of basis functions $B_\alpha$:

$$E_m^{MLP} = \sum_{i=1}^{V} V_m(\boldsymbol{n}_i) \ , \tag{10}$$

$$V(\boldsymbol{n}_i) = \sum_{\alpha} \xi_\alpha B_\alpha(\boldsymbol{n}_i) \,, \tag{11}$$

where the basic function $B_\alpha$ is constructed based on $i_{\text{th}}$ atomic environment $\boldsymbol{n}_i$ via moment tensor descriptor and $\xi_\alpha$ describes the minimized difference between MLIP and AIMD energies.

In machine learning networks, the Taylor expansion of the potential energy $E^{MLP}$ of the configuration at small displacements can be obtained:

$$E^{MLP} = E_0^{MLP} + \frac{1}{2}\sum_{\substack{ij\\ \alpha\beta}} \Phi_{ij}^{\alpha\beta} r_i^\alpha r_j^\beta + \frac{1}{3!}\sum_{\substack{ijk\\ \alpha\beta\gamma}} \Phi_{ijk}^{\alpha\beta\gamma} r_i^\alpha r_j^\beta r_k^\gamma + \frac{1}{4!}\sum_{\substack{ijkl\\ \alpha\beta\gamma\delta}} \Phi_{ijkl}^{\alpha\beta\gamma\delta} r_i^\alpha r_j^\beta r_k^\gamma r_l^\delta + \cdots, \tag{12}$$

where the coefficients $\Phi_{ij}^{\alpha\beta} r_i^\alpha r_j^\beta$, $\Phi_{ijk}^{\alpha\beta\gamma} r_i^\alpha r_j^\beta r_k^\gamma$, $\Phi_{ijkl}^{\alpha\beta\gamma\delta} r_i^\alpha r_j^\beta r_k^\gamma r_l^\delta$ correspond to the second-order, third-order, and fourth-order force constants, respectively. Further, the force constants can be obtained by the finite displacement difference method through our recompiled *thirdorder_MTP.py* and *Fourthorder_MTP.py* code, which provide an interface between MLIP and shengBTE. In fact, the same holds true for other potential functions [66,67]. Finally, the thermal conductivity is obtained by implementing shengBTE with the force constant as input to solve the phonon Boltzmann transport equation: [38,65]

$$\kappa = \kappa_{\alpha\alpha} = \frac{1}{V}\sum_\lambda C_\lambda v_{\lambda\alpha}^2 \tau_{\lambda\alpha} \,, \tag{13}$$

where $\tau_{\lambda\alpha}$, $C_\lambda$, and $v_{\lambda\alpha}$, and $V$ are relaxation time, specific heat capacity, group velocity for $\lambda$ mode phonon along $\alpha$ direction, and crystal volume, respectively. More information about the phonon BTE can be found in Supplementary Note S2.

## ACKNOWLEDGEMENTS

This work is supported by the National Natural Science Foundation of China (Grant No. 52006057), the Natural Science Foundation of Chongqing, China (No. CSTB2022NSCQ-MSX0332), the Fundamental Research Funds for the Central Universities (Grant Nos. 531119200237), and the State Key Laboratory of Advanced Design and Manufacturing for Vehicle Body at Hunan University (Grant No. 52175013). H.W. is supported by the National Natural Science Foundation of China (Grant No. 51906097). Z.Q. is supported by the National Natural Science Foundation of China (Grant No.12274374, 11904324) and the China Postdoctoral Science Foundation (2018M642776). The numerical calculations in this paper have been done on the supercomputing system of the E.T. Cluster and the National Supercomputing Center in Changsha.

## AUTHOR CONTRIBUTIONS

*G.Q.* supervised the project. *L.Y.* performed all the calculations, analysis and writing. *K.D.* performed the writing and analysis. All the authors contributed to interpreting the results. The manuscript was written by *L.Y.* and *K.D.* with contributions from all the authors.

**COMPETING INTERESTS**

The Authors declare no Competing Financial or Non-Financial Interests

**DATA AVAILABILITY**

The data that support the findings of this study are available from the corresponding author on reasonable request.